\documentclass[aps,prd,twocolumn,nofootinbib]{revtex4}
\usepackage{graphicx}
\usepackage{amsfonts,amsmath,amssymb,bm}

\def\be{\begin{equation}}
\def\te{\end{equation}}
\def\ee{\end{equation}}
\def\ba{\begin{eqnarray}}
\def\bea{\begin{eqnarray}}

\def\tea{\end{eqnarray}}
\def\ea{\end{eqnarray}}
\def\eea{\end{eqnarray}}

\begin{document}


\title{Effective dynamics of a nonabelian plasma out of equilibrium}

\author{J. Peralta-Ramos}
\email{jperalta@df.uba.ar}
\affiliation{Departamento de F\'isica, Facultad de Ciencias Exactas y Naturales, Universidad de Buenos Aires and IFIBA, CONICET, Cuidad Universitaria, Buenos Aires 1428, Argentina}

\author{E. Calzetta}
\email{calzetta@df.uba.ar}
\affiliation{Departamento de F\'isica, Facultad de Ciencias Exactas y Naturales, Universidad de Buenos Aires and IFIBA, CONICET, Cuidad Universitaria, Buenos Aires 1428, Argentina}

\begin{abstract}
Starting from kinetic theory, we obtain a nonlinear dissipative formalism describing the nonequilibrium evolution of scalar colored particles coupled selfconsistently to nonabelian classical gauge fields. The link between the one-particle distribution function of the kinetic description and the variables of the effective theory is determined by extremizing the entropy production. This method does not rely on the usual gradient expansion in fluid dynamic variables, and therefore the resulting effective theory can handle situations where these gradients (and hence the  momentum-space anisotropies) are expected to be large.  The formalism presented here, being computationally less demanding than kinetic theory, may be useful as a simplified model of the dynamics of color fields during the early stages of heavy ion collisions and in phenomena related to parton energy loss.

\end{abstract}

\maketitle
\newpage 


\section{Introduction}

The results of numerous experiments on ultrarelativistic heavy ion collisions carried out at the Relativistic Heavy Ion Collider (RHIC) and the Large Hadron Collider (LHC) clearly point to the conclusion that a hot, dense and opaque nuclear medium is created in such events \cite{RHIC-disc1,RHIC-disc2,RHIC-disc3,RHIC-disc4,RHIC-disc5,RHIC-disc6}. By now, the standard picture for the evolution of such matter consists in a pre-equilibrium stage dominated by strong chromoelectromagnetic fields, followed by the formation of the Quark Gluon Plasma (QGP) in local thermal equilibrium. The QGP thus formed expands and cools under its own pressure, going through the deconfinement transition in which hadrons are formed and later on detected (for reviews see \cite{revhydro1,revhydro2,reviewIANCU} and references therein).  

Before the collision, the occupation number of gluons at
high energy is so large that  an approximation in terms of classical gauge 
fields obeying Yang-Mills equations becomes more suitable than a description in terms of on-shell particles \cite{reviewIANCU,GLASMA}. The time scale for this highly nonlinear regime occurring at the earliest stage of a heavy ion collision is $Q_s^{-1}\sim 0.2$ fm/c at RHIC, where $Q_s$ is the saturation scale. 

The dense system appearing between first impact and the formation of the equilibrated QGP develops Chromo-Weibel instabilities (see e.g. \cite{strick-Weibel,arn-Weibel}): the non-equilibrium anisotropic distribution of
the partons is responsible for the fast growth of the chromomagnetic plasma modes, which in turn isotropize the system and speed up the thermalization process to yield a thermalization time $\sim 1$ fm/c. This is roughly the value of the thermalization time that can be inferred from a comparison of state-of-the-art hydrodynamic simulations to data. Moreover, a fast parton transversing the matter formed in heavy ion collisions excites color fields and loses energy, a phenomenon known  as jet quenching \cite{loss1,loss2}. For these reasons, the dynamics of classical gauge fields (both coupled or not to hard partons) has received a lot of attention in the context of heavy ion phenomenology  \cite{reviewIANCU,GLASMA,romstrick,mro93,mro94,mro97,Selikhov,nayak,GLASMA,muller96therm,muller96lattice,bergesgauge,DuslingPhi,florMHD,Gale,fujii,strick-Weibel,arn-Weibel,mannakin,fluct,strick,flor,dum,kurk,mromull,mullerloss,losscomp,losscomp2,rom1,rom2,carrdiel,mro89,sch,sch-WYM,muller-pos,Tani,ipp,stein,ManMro,manna,jiang,holm,wakejiang,white,localeq}.

The color fields ``see" the evolving {\it matter} (which in our case is composed of colored partons) through the conserved color current, as dictated by Yang-Mills equations. The dynamics of the conserved currents, i.e. the color current and the total energy-momentum tensor of the combined system of matter plus gauge fields, is in general very complex, since in principle it must be computed from the dynamics of the matter fields (or particles in the kinetic limit). 

In the semiclassical kinetic approach, the coupling of hard and soft modes is implemented through a nonabelian transport equation \cite{Heinz,HeinzPRL} which determines the evolution of a one-particle distribution function $\mathbf{f}$, which is a matrix in color space (see Section \ref{kin}), or equivalently through Wong's equations \cite{wong}; see also Refs. \cite{libro,elze,Litim,winter,revBlaizot,BlaizotIancu, BraPis90a,FreTay90,Bod98,Bod99,ArSoYa99a,ArSoYa99b}. Once $\mathbf{f}$ is known, the color current acting as the source in Yang-Mills equation is completely determined, so the dynamics of the gauge fields can be found. 
 
Here, we take the view that it is a reasonable hypothesis to assume that the dynamics of the conserved currents is {\it largely} determined by the conservation laws themselves. This opens up the possibility of constructing an effective theory incorporating the conservation laws, that allows one to investigate relevant aspects of the dynamics of gauge fields (for example, the border between stability and instability, or the back reaction of hard particles on the evolution of color fields) in a simpler context as compared to the microscopic theory. 

The degrees of freedom of the effective theory include hydrodynamic variables such as flow velocity. However, since we are working in a strongly out of equilibrium regime, it is not possible to obtain a closed dynamics based on hydrodynamic variables alone.


We obtain a closed theory by identifying a tensor $\lambda^{\mu\nu}$ (which is introduced in Section \ref{eff}) through which the system couples to the hydrodynamic variables. One can think of the introduction of the nonhydrodynamic variable $\lambda^{\mu\nu}$ as a simple way of modeling the back reaction of the distribution function (which could correspond to a highly nonequilibrium situation) on the hydrodynamic modes (i.e. those modes associated to conservation laws and hence relaxing much more slowly). A somewhat similar situation is discussed in \cite{tanos,tanos2,tanos3,anile,muscato} in the context of the Entropy Maximum Principle (EMP), in \cite{GKpaper} (see also \cite{GKbook}) where the moments of the collision term of the classical Boltzmann equation are interpreted as independent variables rather than as infinite moment series, in \cite{mauricio} and \cite{aniso} in the context of ``anisotropic hydrodynamics'', in \cite{nagy} in the framework of divergence-type theories \cite{geroch} (see also \cite{dev,app,linking}), in \cite{denicol} where the closure is obtained from an expansion of the distribution function in moments {\it at all orders}, and in \cite{christen,christen2} in relation to the Entropy Production Principle (EPP). 

Here, the tensor $\lambda^{\mu\nu}$ is identified with a Lagrange multiplier in a well-defined variational problem whose solution yields the distribution function that extremizes the entropy production (for a review of this method see \cite{epvm}). This procedure results in a closure for the distribution function of colored particles which has two satisfying properties: (i) it is nonlinear in the variables of the effective theory, thus generalizing Grad's quadratic ansatz \cite{deGroot,ferz,is1,liboff,kremer} in a nontrivial way \cite{linking,progress}, and (ii) it does not rely in any way on the gradient expansion in fluid variables. The equation of motion for $\lambda^{\mu\nu}$ is then obtained from this closure by the method of moments. 

The result is an effective theory for the dynamics of color fields coupled to colored particles that can handle highly nonequilibrium situations, for example the large momentum anisotropy present at early times in heavy ion collisions. Our formalism is a simplified model that shares with the true dynamics the conservation laws, Lorentz and gauge invariance, being causal, and satisfying the Second Law. 

In the spirit of finding the simplest possible theory incorporating the conservation laws, we postulate a simple BGK form for the collision term \cite{liboff,kremer} in the Boltzmann equation, and obtain solutions at second order in the relaxation time. By going to second order in the relaxation time (instead of the usual first order treatment) we expect to broaden even more the range of applicability of the effective theory, to be able to describe better those situations with large momentum-space anisotropies. 

The limitations of the model put forward in this work are discussed in the last section, but it is worth mentioning them here. The model is phenomenological, it is valid only when the gauge fields can be treated classically, and only when the relaxation time approximation is valid. Moreover, the collision term is linear. 

We would like to stress that our approach is not new. The idea of obtaining a closure from a variational principle whose Lagrange multipliers are identified with macroscopic variables is at the heart of the so-called Extended Thermodynamic Theories (see e.g. \cite{extended}), which use as a variational principle the EMP. However, in most applications of this theory the EMP is only used to express $f$ in terms of the Lagrange multipliers {\it but not} to obtain their dynamics. In Refs. \cite{tanos,anile,tanos2,tanos3,muscato} the EMP was used to accomplish both tasks in the context of electron transport through mesoscopic semiconductors in the nonlinear and nonequilibrium regime (i.e. under conditions of very strong electric fields and large gradients). The result was a closed effective theory which could reproduce well the results of kinetic Monte Carlo simulations. 

In this paper, instead of using the EMP we rely on the EPP, which allows us to find the distribution function that extremizes the entropy production subject to constraints on the conserved currents \cite{epvm}. 
We note that the EPP was used in \cite{christen} to construct a model of nonequilibrium  electron transport and in \cite{christen2} to describe radiative heat transport in a photon gas, obtaining results that agree well with more sophisticated approaches. 

The motivation for the choice of the EPP is threefold. This principle is deeply connected to nonequilibrium statistical physics, in particular to the evolution of fluctuations around a stationary state (we shall not discuss this connection here; for details see \cite{epvm,min1,ferz,prigo}). 
Moreover, as it is rigorously shown in Ref. \cite{epvm} for nonrelativistic systems (see also \cite{ferz}), the distribution function obtained from the EPP actually solves the {\it linearized} Boltzmann equation, which means that the closure provided by this variational principle is able, at least in principle, to capture some features of the microscopic dynamics. 
Finally, the transport coefficients obtained from the EMP can differ substantially from those computed from the EPP, the latter providing better agreement with the results obtained from kinetic theory (see for instance \cite{epvm,tanos,anile,tanos2,tanos3,muscato,christen2,app}). 

In relation to our developments, we note in particular Ref. \cite{florMHD} in which the so-called ``anisotropic hydrodynamics'' (that was developed in \cite{aniso,mauricio}) is coupled to color fields in a way reminiscent of magnetohydrodynamics. In \cite{flor}, an effective model for the dynamics of color fields coupled to particles is obtained from the Boltzmann-Vlasov equations by the method of moments and then applied to study chromoelectric oscillations in a dynamically evolving anisotropic background; see also \cite{strick}. 
As it will become clear in what follows, our approach resembles those adopted in these studies in that they are effective theories capable of dealing with large deviations from equilibrium. 

The colorless version of the effective theory developed in this paper was investigated in \cite{dev} (although there it was obtained in a different way as the one followed here). It was then applied to study the evolution of matter created in heavy ion collisions at RHIC and to the calculation of hadronic observables in \cite{app}, and compared to second order fluid dynamics \cite{hyd1,hyd2}, to which the colorless effective theory reduces when deviations from equilibrium are small. The connection between the (colorless) effective formalism and kinetic theory was established in \cite{linking}. 
 
Similarly to what happens in the colorless version of the effective theory \cite{dev,app}, the formalism presented here reduces to the so-called  ``chromohydrodynamics'' \cite{Heinz,ManMro,jiang,holm} when deviations from equilibrium are small (see Section \ref{matrix}). 
Previous studies based on chromohydrodynamics include the calculation of the wake potential induced by a fast parton \cite{wakejiang} as well as collective excitations and instabilities in the QGP \cite{ManMro,manna,jiang,fraga,cse}.

This paper is organized as follows. In Section \ref{setup}, we describe the basic theoretical setup for our developments, give a very brief overview of the kinetic theory of a nonabelian plasma, and introduce the conserved and entropy currents. In Section \ref{eff} we obtain a closure for the distribution function from the entropy production principle and derive the evolution equations of the effective theory by the method of moments applied to the transport equation; this Section contains our main results. In Section \ref{matrix} we compare our developments to the matrix approach to chromohydrodynamics, and as a simple  illustrative example we compute the polarization tensor of the colored plasma including a finite relaxation time for the fluctuations. We conclude in Section \ref{summ} with some comments on the possible  application of the developed formalism to the dynamics of color fields and instabilities in heavy ion collisions. 

\section{Theoretical setup}
\label{setup}

\subsection{The system}
We are interested in obtaining an effective theory for the dynamics of a system of colored particles interacting with nonabelian classical gauge fields. In this work we will deal with scalar particles coming in three colors, which in our simple model would represent massless and spinless quarks. 
We shall therefore consider a classical Yang-Mills field coupled to conformal scalar matter in the fundamental representation of $SU(3)$. 

In what follows, we will use $(\mu,\nu,\ldots)$ to denote world indices and $(a,b,\ldots)$ to denote internal (color) indices. We shall denote with $N=3$ the dimension of the fundamental representation, and use $n$ to indicate a generic dimension ($n=3$ or $n=8$ for the fundamental or adjoint representations, respectively). 
The generators $\mathbf{T}_a$ are traceless hermitian $n\times n$ matrices with commutation relations

\be
\left[\mathbf{T}_a,\mathbf{T}_b\right] =iC^c_{ab}\mathbf{T}_c
\label{0}
\te
and trace

\be
\mathrm{tr}\mathbf{T}_a\mathbf{T}_b=\frac12\delta_{ab}
\label{01}
\te
The Yang-Mills field is $\mathbf{A}_{\mu}=A^a_{\mu}\mathbf{T}_a$. The field tensor 

\be
\mathbf{F}_{\mu\nu}=\partial_{\mu}\mathbf{A}_{\nu}-\partial_{\nu}\mathbf{A}_{\mu}-ig\left[\mathbf{A}_{\mu},\mathbf{A}_{\nu}\right]
\label{1}
\te
belongs to the adjoint representation of the gauge group. 
The equations of motion for the Yang-Mills field are 

\be
\mathbf{D}_{\mu}\mathbf{F}^{\mu\nu}=-\mathbf{Q}\left[\mathbf{J}_{\nu}\right]
\label{3}
\te
where the covariant derivative is

\be
\mathbf{D}_{\mu}\mathbf{X}=\partial_{\mu}\mathbf{X}-ig\left[\mathbf{A}_{\mu},\mathbf{X}\right]
\label{4}
\te
$\mathbf{Q}$ is a projection operator

\be
\mathbf{Q}\left[\mathbf{X}\right]=2\sum_a\mathbf{T}_a\mathrm{tr}\mathbf{T}_a\mathbf{X}=\mathbf{X}-\frac1n\mathrm{tr}\mathbf{X}
\label{4.01}
\te
Eq. (\ref{3}) implies the Bianchi identity

\be
\mathbf{Q}\left[\mathbf{D}_{\mu}\mathbf{J}^{\mu}\right]=0
\label{5}
\te

We also have the energy-momentum tensor

\be
T^{\mu\nu}_{YM}=\mathrm{tr}\mathbf{T}^{\mu\nu}_{YM}
\label{6}
\te
where

\be
\mathbf{T}^{\mu\nu}_{YM}=\mathbf{F}^{\mu}_{\lambda}\mathbf{F}^{\nu\lambda}-\frac14g^{\mu\nu}\mathbf{F}^{\lambda\rho}\mathbf{F}_{\lambda\rho}
\label{7}
\te
is traceless in world indices. Using the identity $\mathbf{D}_{\left(\mu\right.}\mathbf{F}_{\left.\nu\lambda\right)}=0$ (where brackets mean symmetrization) we get

\be
\mathrm{tr}\left\{\mathbf{D}_{\mu}\mathbf{T}^{\mu\nu}_{YM}+\mathbf{J}_{\lambda}\mathbf{F}^{\nu\lambda}\right\}=0
\label{8}
\te

\subsection{Kinetic theory}
\label{kin}

In principle the scalar matter should be described by quantum field theory. The reduction of the nonequilibrium quantum field description to kinetic theory is fairly established by now, see e.g. \cite{libro} and references therein.

The kinetic equation which governs the evolution of the one-particle distribution matrix $\mathbf{f}$ reads \cite{HeinzPRL,Heinz} (see also \cite{libro,elze,BlaizotIancu,winter,Litim})
\be
p^{\mu}\left[\mathbf{D}_{\mu}\mathbf{f}-\frac g2\left(\mathbf{F}_{\mu\nu}\frac{\partial\mathbf{f}}{\partial p_{\nu}}+\frac{\partial\mathbf{f}}{\partial p_{\nu}}\mathbf{F}_{\mu\nu}\right)\right]=\mathrm{sign}\left(p^0\right)\mathbf{I}_{col}
\label{9}
\te
where 
\be 
\mathbf{D}_\mu \mathbf{f} = \partial_\mu \mathbf{f} - ig [\mathbf{A}_\mu,\mathbf{f}]
\te
with $\mathbf{A}_\mu$ expressed in the fundamental representation. $\mathbf{f}\left(X,p\right)$  is an $N\times N$ matrix ($N=3$ for quarks) and obeys $\mathbf{f}^{\dagger}\left(X,p\right)=\mathbf{f}\left(X,p\right)$.
   
The collision kernel on the right hand side encodes the interaction among the hard excitations of the matter field, and is, in general, a complicated functional of the self-energy \cite{libro}. However, to carry out our developments we really do not need to consider it in much detail, and we anticipate that later on we will use a phenomenological linear collision operator that will suffice for our present purposes.


\subsection{Conserved currents}
\label{currents}

We now introduce the matter entropy and the conserved currents of the microscopic theory. The nonabelian current reads
\be
\mathbf{J}_{\lambda}=g\int\:Dp\;p_{\lambda}\mathbf{f}
\label{10}
\te
where $Dp=d^4p\delta\left(p^2\right)/\left(2\pi\right)^3$.

We have not written down an explicit equation for the matter stress-energy tensor $T^{\mu\nu}_{m}$, but we know that since the total stress-energy must be conserved, we must have 
\be 
T^{\mu\nu}_{m;\nu}=-T^{\mu\nu}_{YM;\nu}=
\mathrm{tr}\mathbf{J}_{\lambda}\mathbf{F}^{\mu\lambda}
\label{ecT}
\te 
We get this by writing
\be
T^{\mu\nu}_{m}=\mathrm{tr}\mathbf{T}^{\mu\nu}_{m}
\label{11}
\te

\be
\mathbf{T}^{\mu\nu}_{m}=\int\:Dp\;p^{\mu}p^{\nu}\mathbf{f}
\label{12}
\te
In Eq. (\ref{ecT}), the semicolon stands for an ordinary derivative. We shall drop the subindex $m$ for $\mathbf{T}^{\mu\nu}_{m}$ and write $\mathbf{T}^{\mu\nu}$ in what follows.

Eqs. (\ref{5}) and (\ref{8}) are identically satisfied provided

\be
\begin{split}
&\int\:Dp\:\mathrm{sign}\left(p^0\right)\mathrm{tr}\left(\mathbf{T}_a\mathbf{I}_{col}\right)=\\
&\mathrm{tr}\int\:Dp\:\mathrm{sign}\left(p^0\right)p^{\mu}\mathbf{I}_{col}=0
\label{13}
\end{split}
\te
The entropy current is 

\be
S^{\mu}=\int\:Dp\;p^{\mu}\mathrm{sign}\left(p^0\right)\mathrm{tr}\left\{\left(\mathbf{1}+\mathbf{f}\right)\ln\left(\mathbf{1}+\mathbf{f}\right)-\mathbf{f}\ln\mathbf{f}\right\}
\label{14}
\te
leading to the entropy production

\be
S^{\mu}_{;\mu}=\int\:Dp\;\mathrm{tr}\left\{\mathbf{I}_{col}\ln\mathbf{f}^{-1}\left(\mathbf{1}+\mathbf{f}\right)\right\}
\label{15b}
\te

We note that to go from Eq. (\ref{14}) to Eq. (\ref{15b}) one must assume that $[\mathbf{f},\mathbf{D}_\mu \mathbf{f}]=0$ (see \cite{localeq}). If this condition is not imposed on the distribution function, the entropy production contains terms of the form $\rm{tr} \{\mathbf{F}_{\mu\nu},\mathbf{f}\}$ which contribute to entropy production even in mean field. Assuming $[\mathbf{f},\mathbf{D}_\mu \mathbf{f}]=0$ then corresponds to assuming that entropy is produced solely by collisions among the particles. We shall stick to this approximation in what follows. 

\section{Effective theory}
\label{eff}

\subsection{Obtaining a closure for $\mathbf{f}$}
\label{closure}

We will now obtain an expression for the one-particle distribution matrix in terms of variables of the effective theory. To do so, we will rely on the EPP \cite{epvm} discussed in the Introduction. 

This method allows one to find the distribution function which extremizes the entropy production $S^\mu_{;\mu}$, subject to the constraints that the conserved currents take on known values. It provides a prescription to associate a distribution function to given macroscopic currents, yielding a nonlinear closure that generalizes the well-known Grad's quadratic ansatz \cite{deGroot,is1,ferz,liboff,kremer} in a nontrivial way \cite{linking,progress}. 

The EPP does not rely on a gradient expansion, which is usually invoked when deriving hydrodynamics from kinetic theory. This results in effective theories capable of describing highly nonequilibrium and nonlinear situations quite reliably as compared to microscopic approaches \cite{epvm,tanos,anile,tanos2,tanos3,muscato,christen,christen2}. 


\subsubsection{Deviations from the unperturbed state}

For simplicity, in what follows we will neglect quantum statistics. 
Given $\rm{tr}(\mathbf{T}^{\mu\nu})=T^{\mu\nu}$, we can define a flow velocity $u^\mu$ and a temperature $T$ by using the Landau-Lifshitz prescription. We have 
\be 
u_\mu T^{\mu\nu}  = \rho(T) u^\nu
\te 
where $\rho(T)$ is the energy density as obtained from the equation of state. 
The prescription amounts to matching the local nonequilibrium state of the flowing real matter to a fiducial perfect fluid. We will show later that the stress tensor $\Pi^{\mu\nu}$ is transverse, i.e. $u_\mu\Pi^{\mu\nu}=0$, so the prescription is consistent.
With $u^\mu$ and $T$, we can define $\beta^\mu=u^\mu/T$ and then construct 
\be 
f_0(x^\mu,p^\mu) = e^{-\beta_\nu p^\nu}
\te

We write the distribution function as 
\be 
\mathbf{f}= f_0 [\mathbf{1}+(1+f_0)\mathbf{\chi}] \approx f_0 [\mathbf{1}+\mathbf{\chi}]
\label{fandchi}
\te 
The entropy production 
\be 
S^{\mu}_{;\mu}=-\int\:Dp\;\mathrm{tr}\left\{\mathbf{I}_{col}\ln\mathbf{f}\right\}
\te 
then reads 
\be 
S^{\mu}_{;\mu}=-\int\:Dp\;\mathrm{tr}\left\{\mathbf{I}_{col}\ln(\mathbf{1}+\mathbf{\chi})\right\}
\label{smumubis}
\te 
where we have used that 
\be 
\int\:Dp\;\mathrm{tr}\left\{\mathbf{I}_{col}\ln(f_0)\right\} = 0
\te

Given the expression for the entropy production given by Eq. (\ref{smumubis}) and taking into account that we will consider a linear collision operator, in order to satisfy the H-theorem we introduce a new variable $\mathbf{Z}$ such that 
\be 
e^{\mathbf{Z}} = \mathbf{1} + \mathbf{\chi}
\te 
The solution to the variational problem entails finding $\mathbf{Z}$ as a function of the Lagrange multipliers to be introduced shortly. The outcome is a closure for the distribution function $\mathbf{f}$, i.e., an expression for $\mathbf{f}$ in terms of the variables of the effective theory, which are the usual hydrodynamic variables and the Lagrange multipliers. The latter encode the back reaction of the distribution function on the hydrodynamic modes. 

The basic plan we will follow is to divide relevant variables into colorless and colored pieces. Therefore, we parametrize $\mathbf{Z}$ as follows 
\be 
\mathbf{Z} = \frac{1}{N}\zeta \mathbf{1} + \zeta^a \mathbf{T}^a
\label{parz}
\te 
The quantities $\zeta^a$ can be identified with color fugacities $\zeta^a \equiv \mu^a/T$, where $\mu^a$ are the color chemical potentials needed to conserve color (see Eq. (\ref{13})).  We shall show later that the color chemical potentials must adjust to the flow of matter as well as to the evolving gauge fields in order to make the system globally colorless, resulting in a highly nontrivial dynamics for the system of colored particles and gauge fields. 

We will work to quadratic order in $\mathbf{Z}$, so we get 

\be 
\mathbf{\chi} = \mathbf{Z} + \frac{1}{2}\mathbf{Z}^2 
\te 
Using that 
\be 
\mathbf{T}^a\mathbf{T}^b = \frac{1}{2N} \delta^{ab}\mathbf{1} + K^{ab}_c \mathbf{T}^c
\te
with 
\be
  K^{ab}_c \equiv \frac{1}{2}(i C_{abc}+ d_{abc})
\te 
where $d_{abc}$ are the symmetric structure constants, 
we get 
\be 
\mathbf{Z}^2 = \frac{1}{N^2}\zeta^2 \mathbf{1} + \frac{2}{N} \zeta \zeta^a \mathbf{T}^a 
+ \frac{1}{2N}\zeta^a \zeta^a \mathbf{1} + \frac{1}{2}\zeta^a \zeta^b d^{ab}_c \mathbf{T}^c
\te

Having $\mathbf{Z}$, we obtain 

\be 
\begin{split}
\mathbf{\chi} &= \frac{1}{N}\zeta \mathbf{1} + \zeta^a \mathbf{T}^a + \frac{1}{2}\bigg[
\frac{1}{N^2} \zeta^2 \mathbf{1} \\
& + \frac{2}{N}\zeta\zeta^a  \mathbf{T}^a + 
\frac{1}{2N}\zeta^a \zeta^a \mathbf{1} + \frac{1}{2}\zeta^a \zeta^b d^{ab}_c \mathbf{T}^c \bigg] 
\end{split}
\te

\subsubsection{Currents, entropy production and collision term}

To solve the variational problem we must express the currents and the entropy production in terms of $\mathbf{Z}$. 

The shear tensor 
\be 
\Pi^{\mu\nu} = \mathrm{tr} \left\langle p^\mu p^\nu \mathbf{\chi}  \right\rangle 
\te 
reads
\be 
\Pi^{\mu\nu} = \left\langle p^\mu p^\nu  \bigg[ \zeta + \frac{1}{2}\bigg( \frac{1}{N}\zeta^2 
+ \frac{1}{2}\zeta^a \zeta^a  \bigg) \bigg]  \right\rangle
\label{shearfull}
\te 
where we have introduced the notation 
\be 
\int Dp f_0 (\ldots) \equiv \left\langle (\ldots)  \right\rangle 
\te

From Eq. (\ref{10}) we get the expression for the color currents
\be 
\mathbf{J}^\mu = \bar{n}u^\mu \bigg[\zeta^a \mathbf{T}^a + 
\frac{1}{4}d^{ab}_c \zeta^a \zeta^b \mathbf{T}^c  \bigg] + \frac{g}{N}\left\langle p^\mu \zeta \right\rangle \zeta^a \mathbf{T}^a 
\label{jdfull}
\te
where we have defined $\bar{n}=g\left\langle \omega \right\rangle$. 

For the entropy production, we have

\be 
S^{\mu}_{;\mu}= -\int Dp ~\mathrm{tr} (\mathbf{I}_{col} \mathbf{Z})
\te

The collision operator contains color independent and dependent parts, so, similarly to the decomposition used for $\mathbf{Z}$, we put 
\be 
\mathbf{I}_{col} = I^{(0)}_{col} \mathbf{1} + I^{a}_{col} \mathbf{T}^{a}
\te 
We then have 
\be 
 \mathrm{tr} (\mathbf{I}_{col} \mathbf{Z}) = I^{(0)}_{col}\zeta + \frac{1}{2}I^{a}_{col}\zeta^a
\te 
where we have used Eq. (\ref{01}), so
\be 
S^{\mu}_{;\mu}= - \int Dp~ \bigg(I^{(0)}_{col}\zeta + \frac{1}{2}I^{a}_{col}\zeta^a \bigg)
\label{entaux}
\te 

We shall write the collision operator as 
\be 
\mathbf{I}_{\beta} = -\frac{1}{2\tau} R[FR[\mathbf{Z}]]
\label{linea}
\te
where we have put 
\be 
\mathbf{I}_{col} \equiv f_0 \mathbf{I}_{\beta} 
\te 

In Eq. (\ref{linea}), $\tau$ is the relaxation time, $F=F(\omega)$ is an arbitrary function of energy $\omega=-u_\mu p^\mu$, and $R$ is an operator enforcing the integrability conditions given in Eqs. (\ref{13}). The projector $R$ explicitly reads 
\be 
R[\mathbf{Z}] = \mathbf{Z} - R^{ab}\left\langle \mathrm{tr}(\mathbf{T}_a \mathbf{Z}) \mathbf{T}_b \right\rangle 
- \mathbf{1} R_\mu \left\langle p^\mu \mathrm{tr}(\mathbf{Z})\right\rangle 
\label{Rexplicit}
\te 
with 
\be 
R_\mu=\frac{1}{N\left\langle \omega \right\rangle} u_\mu + \frac{1}{N\left\langle \omega^2 \right\rangle} \Delta^\nu_\mu p_\nu
\te
and $R_{ab}=2\delta_{ab}$. Here, 
\be
\Delta^{\mu\nu} = g^{\mu\nu}+u^\mu u^\nu
\te 
is the spatial projector. 
Note that this form for $\mathbf{I}_{\beta}$ guarantees that the Second Law holds exactly. Moreover, it is flexible enough to include the important cases of Marle's relativistic generalization of the BGK model \cite{liboff,kremer}, corresponding to $F(\omega)=T$, as well as the Anderson-Witting model corresponding to $F(\omega)=\omega$ \cite{kremer} (for a discussion of these models in connection to the freeze out stage in heavy ion collisions see \cite{linking,progress}).

From Eq. (\ref{linea}) we then have 
\be 
I^{(0)}_{\beta} = -\frac{1}{2\tau N } R[FR[\zeta]]
\label{i0aux}
\te 
and 
\be 
I^{a}_{\beta} = -\frac{1}{2\tau} R[FR[\zeta^{a}]]
\label{iaaux}
\te  
with
\be 
R[\zeta] = \frac{1}{N}\zeta - R_\mu \left\langle p^\mu \zeta \right\rangle
\te 
and 
\be 
\mathbf{T}_a R[\zeta_a] = \mathbf{T}_a \zeta_a - \frac{1}{2}R^{bc} \mathbf{T}_c 
\left\langle \zeta_b \right\rangle
\label{ra}
\te 

The integrability conditions then read
\be 
\left\langle p^\rho I^{(0)}_{\beta}  \right\rangle = 0
\label{int0}
\te 
and 
\be 
\left\langle I^{a}_{\beta}  \right\rangle = 0
\label{inta}
\te 
Note that there are no restrictions on $\left\langle p^\rho I^{a}_{\beta}  \right\rangle$ or on $\left\langle I^{(0)}_{\beta}  \right\rangle$. 

\subsubsection{Variational equations}

In our case the conserved currents are $J^{a\mu}$ and $T^{\mu\nu}$, so that the variational problem becomes
\be 
\frac{\delta}{\delta \mathbf{Z}} [S^{\mu}_{;\mu} - \lambda_{\mu\nu}T^{\mu\nu}-\lambda_{a\mu}J^{a\mu}] = 0
\te 
where $\lambda_{\mu\nu}$ and $\lambda_{a\mu}$ are Lagrange multipliers forcing the energy momentum tensor and the color currents to take on their known values.

We then have 
\be 
\begin{split}
 \frac{\delta S^{\mu}_{;\mu}}{\delta \zeta} &= \lambda_{\mu\nu} \frac{\delta T^{\mu\nu}}{\delta \zeta} + \lambda_{a\mu}
 \frac{\delta J^{a\mu}}{\delta \zeta} \\
\frac{\delta S^{\mu}_{;\mu}}{\delta \zeta^{b}} &= \lambda_{\mu\nu} \frac{\delta T^{\mu\nu}}{\delta \zeta^{b}} + \lambda_{a\mu}
 \frac{\delta J^{a\mu}}{\delta \zeta^{b}}
\end{split}
\label{varia}
\te

Using Eqs. (\ref{shearfull}), (\ref{jdfull}) and (\ref{entaux}), together with Eqs. (\ref{i0aux}) and (\ref{iaaux}), the variational equations (\ref{varia}) become 
\be 
\frac{1}{\tau N } R[FR[\zeta]] = \lambda_{\mu\nu} p^\mu p^\nu \bigg[1+\frac{1}{N} \zeta \bigg] + g\lambda_\mu^a p^\mu \bigg[  \frac{1}{N}\zeta^a  \bigg]
\label{variat1}
\te 
and 
\be 
\frac{1}{\tau } R[FR[\zeta^{b}]] = \lambda_{\mu\nu} p^\mu p^\nu \zeta^b 
+ g\lambda_\mu^a p^\mu \bigg[\frac{1}{N}\zeta\delta^{ab} + \frac{1}{2}\zeta^d d^{bd}_a\bigg]
\label{variat2}
\te 

We will solve the equations (\ref{variat1}) and (\ref{variat2}) to second order in the relaxation time $\tau$. To this end, we will follow \cite{linking} and expand the nonequilibrium correction $\mathbf{Z}$ and the Lagrange multipliers as follows
\be 
\begin{split}
\zeta &= \zeta^{(1)} + \zeta^{(2)} \\
\zeta^{a} &= \zeta^{a(1)} + \zeta^{a(2)} \\
\lambda_{\mu\nu} &=  \lambda_{\mu\nu}^{(1)} + \lambda_{\mu\nu}^{(2)}\\
\lambda^a_{\mu} &=  \lambda_{\mu}^{a(1)} + \lambda_{\mu}^{a(2)}
\end{split}
\label{zetas2nd}
\te 

We now go over to solve the equations at first and second order in $\tau$. We note that the solution given here closely follows the one given in \cite{linking}. For the reader's convenience, a brief summary of the logical steps carried out to obtain the closure for the distribution function can be found in Section (\ref{sumclos}).

\subsubsection{First order solution}

At first order, the variational equations are
\be 
\frac{1}{\tau N } R[FR[\zeta^{(1)}]] = \lambda^{(1)}_{\mu\nu} p^\mu p^\nu 
\label{var1}
\te 
and 
\be 
\frac{1}{\tau } R[FR[\zeta^{b(1)}]] = 0
\label{var2}
\te 
Eqs. (\ref{ra}) and (\ref{var2}) imply that $\zeta^{b(1)}$ must be independent of $p^\nu$, as it must be given that $\zeta^{b(1)}$ are color fugacities. 

From Eq. (\ref{var1}) we get 
\be
\lambda^{(1)}_{\mu\nu} \left\langle p^\mu p^\nu p^\rho \right\rangle= 0
\label{l0var}
\te 
Setting $\rho=k$ we obtain $\lambda^{(1)}_{0k}=0$. Without loss of generality we can take $\lambda^{(1)}_{00}=0$. Since we are dealing with a conformal theory $\left\langle p^i p^j \right\rangle = \delta^{ij} \left\langle \omega^2 \right\rangle/3$, and we get $\lambda^{(1)i}_i = 0$. We thus obtain 
\be 
\zeta^{(1)}= \frac{\tau N}{F} \lambda^{(1)}_{ij} p^i p^j
\label{zeta10}
\te 

The first order shear tensor is then 
\be 
\Pi_1^{\mu\nu} = \left\langle p^\mu p^\nu \zeta^{(1)} \right\rangle
\te 
We find that $\Pi^{00}=\Pi^{k0}=0$. Using that 
\be 
 \left\langle G(\omega) p^i p^j p^k p^l \right\rangle = \frac{ \left\langle G(\omega)\omega^4 \right\rangle }{15}(\delta^{ij}\delta^{kl}+
\delta^{ik}\delta^{jl}+\delta^{il}\delta^{jk})
\label{idenG}
\te 
for any function of energy $G(\omega)$, we get
\be 
\Pi_1^{ij} = \frac{2\tau N}{15}  \left\langle \frac{\omega^4}{F}  \right\rangle \lambda^{(1)ij}
\label{pi1aux}
\te
which is traceless and transverse.   
The first order color current is (recall that $\zeta^{a(1)}$ must be independent of $p^\nu$)
\be 
J_1^{a\mu} = \bar{n} \zeta^{a(1)} u^\mu
\label{j1a}
\te

The lowest order nontrivial contributions to the entropy current and the entropy production are 
\be 
S^0_1 = -\frac{\tau^2 N}{15}  \left\langle \frac{\omega^5}{F^2}  \right\rangle \lambda^{(1)ij}\lambda^{(1)}_{ij}  \qquad \rm{and}  \qquad 
S^i_1 = 0
\label{ecur}
\te 
in the rest frame, 
and 
\be 
S^\mu_{;\mu} = \frac{2\tau N}{15}  \left\langle \frac{\omega^4}{F}  \right\rangle \lambda^{(1)ij}\lambda^{(1)}_{ij}
\label{eprod}
\te
respectively. The entropy flux (\ref{ecur}) and the entropy production (\ref{eprod}) computed from the first order solutions $(\zeta^{(1)},\zeta^{a(1)})$ to the variational equations are already quadratic in deviations from equilibrium, so we do not need to consider higher order contributions. Note that, because of the first order equation (\ref{var2}), $\zeta^{a(1)}$ does not contribute to the entropy production at quadratic order.

\subsubsection{Second order solution}
The variational equations at second order read
\be 
\begin{split}
\frac{1}{\tau N } R[FR[\zeta^{(2)}]] &= \lambda^{(2)}_{\mu\nu} p^\mu p^\nu + \frac{1}{N} \lambda^{(1)}_{\mu\nu} p^\mu p^\nu \zeta^{(1)}\\
& + \frac{g}{N} \lambda_\mu^{a(1)}p^\mu  \zeta^{a(1)} 
\end{split}
\te 
and 
\be 
\frac{1}{\tau } R[FR[\zeta^{b(2)}]] = g\frac{1}{N}\lambda_\mu^{b(1)} p^\mu \zeta^{(1)} + \lambda_{\mu\nu}^{(1)} p^\mu p^\nu \zeta^{b(1)} 
\te

The integrability conditions then read
\be
\tau N \lambda_{\mu\nu}^{(2)}\left\langle p^\mu p^\nu \omega \right\rangle + \frac{1}{N} \left\langle (\zeta^{(1)})^2 F  \omega \right\rangle + g\tau \lambda_0^{a(1)}\zeta^{a(1)} \left\langle \omega^2 \right\rangle = 0
\label{s1}
\te 
\be
\tau N \lambda_{0j}^{(2)}\left\langle p^i p^j \omega \right\rangle + g\tau \lambda_j^{a(1)}\zeta^{a(1)} \left\langle p^i p^j \right\rangle = 0
\label{s2}
\te 
corresponding to Eq. (\ref{int0})
and 
\be
\frac{g\tau}{N}\lambda_{0}^{b(1)}\left\langle \omega \zeta^{(1)} \right\rangle = 0
\label{s3}
\te 
corresponding to Eq. (\ref{inta}). The latter equation shows that $\lambda_{0}^{b(1)}=0$. 

Any term in $\lambda_{\mu\nu}^{(2)}$ which is not strictly required by the integrability conditions can be absorbed into $\lambda_{\mu\nu}^{(1)}$, so  there is no loss of generality if we take $\lambda_{00}^{(2)}= 0$. Moreover, from Eq. (\ref{s2}) we see that $\lambda_{0j}^{(2)}=0$ and $\lambda_{j}^{a(1)}=0$ is a solution to the integrability condition. Therefore, we can write  
\be 
\lambda_{ij}^{(2)}=-\Lambda \delta_{ij}
\te 
where we have put 
\be 
\Lambda = \frac{1}{N^2 \tau \left\langle \omega^3 \right\rangle}\bigg[ \left\langle F \omega(\zeta^{(1)})^2\right\rangle \bigg]
\te 

The quadratic equations then read 
\be 
R[FR[\zeta^{(2)}]] = -\tau N \Lambda \omega^2 + \frac{1}{N}(\zeta^{(1)})^2 F 
\label{sec1}
\te
and 
\be 
R[FR[\zeta^{b(2)}]] = -\tau \Lambda \omega^2 \zeta^{b(1)}
\label{secb}
\te 

Note that the left hand side of Eq. (\ref{sec1}) vanishes when integrated against $\omega$ (which means that the equation has a solution) but does not vanish when integrated agaist $\omega/F$. Therefore, the solution to Eq. (\ref{sec1}) is
\be 
\zeta^{(2)} = \frac{1}{N}(\zeta^{(1)})^2 - N\tau \Lambda \frac{\omega^2}{F}  - A\frac{\omega}{F}
\label{zeta20}
\te 
where 
\be 
A= \left\langle \frac{\omega^2}{F}\right\rangle^{-1} \bigg[\frac{1}{N} \left\langle \omega (\zeta^{(1)})^2 \right\rangle - 
N\tau \Lambda \left\langle \frac{\omega^3}{F}\right\rangle  \bigg]
\te 

Similarly, we obtain from Eq. (\ref{secb})
\be 
\zeta^{b(2)} = -\tau \Lambda \frac{\omega^2}{F} \zeta^{b(1)} + \frac{B^b}{F}
\label{zeta2b}
\te 
where 
\be 
B^b = \tau \Lambda \zeta^{b(1)}\left\langle \omega^2 \right\rangle
\te 
Note that $\Lambda \propto (\zeta^{(1)})^2$ is already quadratic in $\tau$, which means that  $\zeta^{b(2)}$ is actually third order and therefore can be neglected. 

We are ready to compute $\Pi_2^{\mu\nu}$ and $J_2^{a\mu}$; see Eqs. (\ref{shearfull}) and (\ref{jdfull}). We have 
\be 
\Pi_2^{\mu\nu} = \left\langle p^\mu p^\nu \bigg[ \frac{3}{2N}(\zeta^{(1)})^2  + \frac{1}{4}\zeta^{a(1)} \zeta^{a(1)} 
- N\tau \Lambda \frac{\omega^2}{F} 
 -A\frac{\omega}{F} \bigg] \right\rangle
\te
Note that $\Pi_2^{0i}=0$, but $\Pi_2^{00}\neq 0$, so $\Pi_2^{\mu\nu}$ is not the true correction to the energy-momentum tensor, whereby the parameter $T$ in our equations is not the true temperature (that would be measured by an observed moving with the local rest frame). To obtain the physical correction to $T^{\mu\nu}$, which we call $\Pi_{2,phys}^{\mu\nu}$, we must perform a temperature shift. Putting $T=T_{phys}-\delta T$ with 
$\delta T d\rho/dT = \Pi_2^{00}$, the true correction reads
\be 
\Pi_{2,phys}^{ij}= \Pi_2^{ij}-\frac{1}{3}\delta^{ij}\Pi_2^{00}
\te 
which is traceless. Taking into account the above we find
\be
\begin{split}
\Pi_{2,phys}^{ij} &= L [\lambda^{(1)i}_k \lambda^{(1)kj}-\frac{1}{3}\delta^{ij}\lambda^{(1)lm}\lambda^{(1)}_{lm}] \\
& +T\left\langle \omega \right\rangle \Delta^{ij}\zeta^{a(1)} \zeta^{a(1)}
\label{piphys}
\end{split}
\te 
with 
\be 
L =  \frac{\tau^2 N}{45} \left\langle \frac{\omega^6}{F^2} \right\rangle  
\te 
To obtain the last term in Eq. (\ref{piphys}) we have used the identity $\left\langle \omega^2 \right\rangle = 2T\left\langle \omega \right\rangle $. Note also that $\Pi_{2,phys}^{ij}$ is transverse. For simplicity,  in what follows we will drop the subindex ``phys'' in $\Pi_{2,phys}^{ij}$. 

The quadratic contribution to the color current can be computed directly from Eqs. (\ref{jdfull}), (\ref{zeta20}) and (\ref{zeta2b}). We get 
\be 
J_{2c}^{\mu} = \frac{\bar{n}}{4}u^\mu d^{ab}_c \zeta^{a(1)} \zeta^{b(1)} 
\label{j2htot}
\te 
where we have used that $\lambda^{(1)j}_j = 0$ so that the last term in Eq. (\ref{jdfull}) drops out.

Using the closure picked out by the EPP, we have completed the task of expressing the currents $\rm{tr}(\mathbf{T}^{\mu\nu})$ and $\mathbf{J}^\mu$ in terms of the variables of the effective theory. 

\subsubsection{Summary of the EPP method}
\label{sumclos}

For clarity, we now briefly summarize the main logical steps followed to obtain the closure for $\mathbf{f}$. 

From a linear transport equation, we set up the variational problem given in Eqs. (\ref{varia}). The solution to these   equations gives the distribution function that extremizes the production of entropy subject to the constraints that the conserved currents take on known values.
Using the expressions for the stress tensor, the color currents and the entropy production given in Eqs.  (\ref{shearfull}), (\ref{jdfull}) and (\ref{entaux}), respectively, together with the integrability conditions for the collision term, Eqs. (\ref{i0aux}) and (\ref{iaaux}), the variational equations become Eqs. (\ref{variat1}) and (\ref{variat2}). 

We then expand the Lagrange multipliers and the nonequilibrium correction $\mathbf{Z}$ in powers of the relaxation time $\tau$ (Eq. (\ref{zetas2nd})), and solve the variational equations to second order in $\tau$.  The result is an expression for the correction $\mathbf{Z}$ in terms of the Lagrange multipliers, which is given by Eqs. (\ref{zeta10}) and (\ref{zeta20}). From these equations, we can express the shear tensor and the color currents in terms of the Lagrange multipliers, obtaining Eqs. (\ref{pi1aux}), (\ref{j1a}), (\ref{piphys}) and (\ref{j2htot}). 

The distribution function then reads
\be 
\begin{split}
\mathbf{f}&=f_0 \bigg[1+\frac{N\tau}{F}p^i p^j \lambda_{ij}^{(1)}\\
&+
\frac{N^2\tau^2}{2F^2}p^i p^j p^l p^m \lambda_{ij}^{(1)}\lambda_{lm}^{(1)}+\frac{1}{4}\zeta^{a(1)}\zeta^{a(1)} \bigg]\mathbf{1} \\
&+ f_0 \bigg[\zeta^{a(1)}+ \frac{\tau}{F}\lambda_{ij}^{(1)}p^i p^j \zeta^{a(1)} + \frac{1}{4}d_{ef}^a \zeta^{e(1)} \zeta^{f(1)} \bigg]\mathbf{T^a}
\end{split}
\label{closfin}
\te 
For convenience, we denote by $f$ and $f^a$ the colorless and colored parts of $\mathbf{f}$:
\be 
\begin{split}
f&= f_0 \bigg[1+\frac{N\tau}{F}p^i p^j \lambda_{ij}^{(1)}\\
&+
\frac{N^2\tau^2}{2F^2}p^i p^j p^l p^m \lambda_{ij}^{(1)}\lambda_{lm}^{(1)}+\frac{1}{4}\zeta^{a(1)}\zeta^{a(1)} \bigg]
\label{fexp}
\end{split}
\te 
\be 
f^a = f_0 (\zeta^{a(1)}+ \frac{\tau}{F}\lambda_{ij}^{(1)}p^i p^j \zeta^{a(1)} + \frac{1}{4}d_{ef}^a \zeta^{e(1)} \zeta^{f(1)})
\label{fa}
\te

As discussed in the Introduction, this method, or very similar ones based on maximizing the entropy, have been used in different contexts to obtain closures for the distribution function, which resulted in effective models whose dynamics compared well with kinetic theory (see e.g. \cite{epvm,prigo} for a broad perspective and \cite{tanos,anile,tanos2,muscato,tanos3,linking,christen,christen2} for specific applications). 

To obtain a dynamical model, we must now find the equation of motion of the Lagrange multipliers.

\subsection{Equations of motion}  
\label{motion}

The EPP described above has provided us with an expression for the distribution function $\mathbf{f}$ in terms of the reduced set of variables of the effective theory: the usual hydrodynamic variables $u^\mu$, $\rho$, the nonhydrodynamic tensor $\lambda_{\mu\nu}^{(1)}$ encoding the back reaction of  $\mathbf{f}$ on the hydrodynamic modes, and the color fugacities $\zeta^{a(1)}$. This expression is given in Eq. (\ref{closfin}).

The equations of motion for $u^\mu$, $\rho$ and $\zeta^{a(1)}$ are the conservation equations for $T^{\mu\nu}$ and $J^{a\mu}$, respectively. However, the conservation laws are not enough to fully determine the dynamics of the system, and an equation governing the evolution of $\Pi^{\mu\nu}$ must be given. Usually, this is done within the gradient expansion for hydrodynamic variables \cite{deGroot,ferz,kremer,revhydro1,revhydro2,hyd1,hyd2}. Here, instead, $\Pi^{\mu\nu}$ is an algebraic function of  $\lambda^{(1)ij}$ and  $\zeta^{a(1)}$, so we must obtain the evolution equation for $\lambda^{(1)ij}$. We will get the latter from the kinetic equation.  

It will prove convenient to express our results in terms of a new relaxation time $\tau_\pi$ (related by a constant to the previously introduced $\tau$)
\be 
\frac{1}{\tau_\pi} = \frac{6}{\tau}\left\langle \frac{\omega^5}{F^2}\right\rangle^{-1} \left\langle \frac{\omega^4}{F} \right\rangle 
\te 
and the shear viscosity
\be 
\eta = \frac{6\tau_\pi N}{15 T} \left\langle \frac{\omega^5}{F^2}\right\rangle^{-1}\left\langle \frac{\omega^4}{F} \right\rangle^2
\te 
which naturally arise in the context of second order fluid dynamics \cite{revhydro1,revhydro2,hyd1,hyd2}.
For simplicity, we also introduce a new variable 
\be 
\gamma^{ij}\equiv \eta T \lambda^{(1)ij}
\te 

We shall deal with the conservation equations first, and then go over to discuss the evolution equation for $\gamma^{ij}$. 

\subsubsection{Conservation equations}

Inserting the expression for $\mathbf{f}$ given by Eq. (\ref{closfin}) into Eqs. (\ref{5}) and (\ref{8}), and using that  $T^{\mu\nu}_{m,\nu}=-T^{\mu\nu}_{YM,\nu}=
\mathrm{tr}\mathbf{J}_{\lambda}\mathbf{F}^{\mu\lambda}$ in the latter, we get equations of motion for the velocity $u^\mu$, the energy density $\rho$ and $\zeta^{a(1)}$. 

For the matter energy-momentum tensor $T^{\mu\nu} = T^{\mu\nu}_0 + \Pi^{\mu\nu}_1+\Pi^{\mu\nu}_2$, where 
\be 
T^{\mu\nu}_0 = \rho (u^\mu u^\nu +\frac{1}{3}\Delta^{\mu\nu})
\label{T0}
\te 
is the perfect-fluid energy-momentum tensor, 
\be 
\Pi^{\mu\nu}_1 = \gamma^{\mu\nu}
\label{pi1}
\te 
and 
\be 
\begin{split}
\Pi^{\mu\nu}_2 &= \frac{L}{\eta^2 T^2} \bigg(\gamma^\mu_\sigma \gamma^{\sigma\nu} + \frac{1}{3}\Delta^{\mu \nu}\gamma^{\sigma\rho}\gamma_{\sigma\rho}  \bigg)  \\
& + T\left\langle \omega \right\rangle \Delta^{\mu \nu}\zeta^{a(1)} \zeta^{a(1)}
\label{pi2}
\end{split}
\te 
we get 
\be 
T^{\mu\nu}_{0;\nu} + \Pi^{\mu\nu}_{1;\nu} +\Pi^{\mu\nu}_{2;\nu} = \frac{1}{2}\bar{n}(\zeta^{e(1)}+\frac{1}{4}d^{e}_{cd}\zeta^{c(1)} \zeta^{d(1)}) u_\lambda F^{e\mu\lambda}
\label{eomtfinal}
\te 
To avoid being cumbersome, we shall not write down the explicit expression for $T^{\mu\nu}_{;\nu}$, but it follows immediately from Eqs. (\ref{T0})-(\ref{pi2}).

The conservation equations for $u^\mu$ and $\rho$ as given in Eq. (\ref{eomtfinal}) involve the time derivative of $\zeta^{a(1)}$ and $\gamma_{\mu\nu}$ (these quantities appear in the LHS of Eq. (\ref{eomtfinal})). 
The evolution equation for $\zeta^{a(1)}$ is obtained from the (covariant) conservation of $\mathbf{J}^\mu$. We get  
\be
\begin{split}
&\bigg[\bar{n}u^\mu \bigg(\zeta^{f(1)} + \frac{1}{4}d^f_{bd}\zeta^{b(1)} \zeta^{d(1)} \bigg) \bigg]_{;\mu}
\\
& + g\bar{n} u^\mu C^f_{ab} A^a_{\mu} \bigg( \zeta^{b(1)} + \frac{1}{4}d^b_{cd}\zeta^{c(1)} \zeta^{d(1)} \bigg) = 0
\label{eomzeta1}
\end{split}
\te 

\subsubsection{Evolution equation}

We shall now deal with the evolution equation for the nonequilibrium tensor $\gamma_{\mu\nu}$. We will obtain this equation from the second moment of the singlet sector of the transport equation Eq. (\ref{9}), i.e. from the equation that results from taking the trace of (\ref{9}). The reason for considering the colorless part of the transport equation will be discussed after presenting the evolution equation.

The second moment of the colorless part of the kinetic equation reads
\be
\begin{split}
&\int\:Dp\; p^\rho p^\sigma p^\mu \bigg[ \rm{tr}(\mathbf{D}_\mu \mathbf{f}) -\frac{g}{2}\rm{tr}\bigg( \mathbf{F}_{\mu\nu}\frac{\partial \mathbf{f}}{\partial p_\nu}+ \frac{\partial \mathbf{f}}{\partial p_\nu}\mathbf{F}_{\mu\nu}\bigg)\bigg] \\
&= \int\:Dp\;{\rm sign}(p^0) p^\rho p^\sigma {\rm tr}(\mathbf{I}_{\rm{col}})
\end{split} 
\label{3moment}
\te

We have that $p^\mu \rm{tr}(\mathbf{D}_\mu \mathbf{f}) = p^\mu\partial_\mu \rm{tr}(\mathbf{f})=p^\mu\partial_\mu f$, where $f$ is the colorless part of $\mathbf{f}$ given in Eq. (\ref{fexp}) at quadratic order.

The second term in the left-hand side of Eq. (\ref{3moment}) becomes
\be
-\frac{g}{2}\int\:Dp\; p^\rho p^\sigma p^\mu F_{\mu\nu}^a \frac{\partial f^a}{\partial p_\nu} 
\te 
where $f^a$ is the colored part of $\mathbf{f}$ given in Eq. (\ref{fa}) at quadratic order.

The right-hand side of Eq. (\ref{3moment}) is 
\be 
N \int\:Dp\; p^\rho p^\sigma  \rm{sign}(p^0) I_{\rm{col}}^{(0)} 
\te 

Using the explicit expressions for $f$ and $f^a$ given in Eqs. (\ref{fexp}) and (\ref{fa}) in the moment equation (\ref{3moment}) we obtain (recall that $\gamma^{ij}\equiv \eta T \lambda^{(1)ij}$ and that, for simplicity, we assume that $\eta$ is a constant)
\be 
\begin{split}
& -\frac{1}{T}M^{\rho\sigma\mu\nu}u_{\nu;\mu} + \frac{N\tau}{\eta T} \bigg( N_1^{ij\rho \sigma \mu} + 
\frac{N\tau}{\eta T} N_2^{ijlm\rho \sigma \mu} \gamma_{lm} \bigg) \gamma_{ij;\mu} \\
&+ \frac{1}{2}N_0^{\rho\sigma\mu} \zeta^{a(1)} \zeta^{a(1)}_{;\mu} 
+ \frac{1}{2} \beta^\nu F^a_{\mu\nu} \bigg( (\zeta^{a(1)}+\frac{1}{4}d^a_{bc}\zeta^{b(1)} \zeta^{c(1)})N_0^{\mu\rho\sigma} \\
&+ \frac{\tau}{\eta T} \zeta^{a(1)} \gamma_{ij} N_1^{\mu\rho\sigma i j}\bigg) 
 -\frac{g\tau}{2\eta T} F^a_{\mu j}\zeta^{a(1)} \gamma_i^{j}N_1^{\mu\rho\sigma i} 
= \\
& -\frac{N}{2\eta T}\gamma_{ij}\bigg( N_1^{ij\rho\sigma} + \frac{\tau}{\eta T} \gamma_{lm} N_2^{ijlm \rho \sigma} \bigg)
\label{eomtemp}
\end{split}
\te 
For brevity, we have defined the following quantities
\be 
\begin{split}
M^{\rho\sigma\mu\nu} &= (1+\frac{1}{4}\zeta^{a(1)} \zeta^{a(1)})N_0^{\rho\sigma\mu\nu} 
- \frac{N\tau}{\eta T}\gamma_{ij} N_1^{\rho\sigma\mu\nu i j} \\
&+ \frac{N^2\tau^2}{2\eta^2 T^2}  \gamma_{ij} \gamma_{lm} N_2^{\rho\sigma\mu\nu i j l m} 
\end{split}
\te 
with 
\be 
 N_\alpha^{\rho\sigma\gamma \cdots \kappa} = \left\langle \frac{1}{F^\alpha} p^\rho p^\sigma p^\gamma \cdots p^\kappa \right\rangle
\label{Ns}
\te 

In order to find an explicit evolution equation for $\gamma_{ij;\mu}$, we must be able to invert the tensor with which it is contracted, namely 
\be 
H^{ij\rho\sigma\mu} \equiv N_1^{ij\rho \sigma \mu} + 
\frac{N\tau}{\eta T} N_2^{ijlm\rho \sigma \mu} \gamma_{lm}
\te 
For our present purposes it is enough to display the equation of motion for the nonequilibrium tensor $\gamma^{ij}$ to linear order. 

Using suitable generalizations of Eq. (\ref{idenG}) to compute the $N's$ explicitly we get (in the local rest frame)

\be 
\begin{split}
\dot{\gamma}^{ij} &= -a_1 \zeta^{a(1)} \Delta^{ij}\dot{\zeta}_1^a -\frac{\eta}{\tau_\pi}(1+\frac{1}{4}\zeta^{a(1)} \zeta^{a(1)})\sigma^{ij} \\
&+ a_2 \zeta^{a(1)} F^{a(i}_\alpha \gamma^{\alpha j)}
-\frac{1}{\tau_\pi}\gamma^{ij} -\bigg[5\frac{\dot{T}}{T} +\frac{1}{3}u^k_{;k} \bigg]\gamma^{ij} \\
&+ \gamma^{i}_k \sigma^{kj}+\gamma^{j}_k \sigma^{ki} -\frac{2}{3}\delta^{ij}\gamma_{kl}\sigma^{kl}
\end{split}
\label{finaleom}
\te 

In Eq. (\ref{finaleom}), $\sigma^{\rho\sigma}$ is the first order shear tensor
\be 
\sigma^{\rho\sigma} = \nabla^{<\rho}u^{\sigma>} 
\te 
where $C^{<\mu\nu>}$ denotes taking the traceless and transverse part of a tensor $C$ and $\nabla^{\mu}=\Delta^{\mu\nu}\partial_\nu$ is the spatial gradient. We have denoted the convective derivative by an overdot, i.e. $\dot{C}=u^\alpha \partial_\alpha C$. The parenthesis around indices denote symmetrization. 

The transport coefficients $a_k$ that appear in the evolution equation are  
\be 
\begin{split}
a_{1}&=\frac{5\eta T}{6N\tau_\pi}\left\langle \frac{\omega^5}{F^2} \right\rangle \left\langle \frac{\omega^5}{F} \right\rangle^{-1}
\left\langle \frac{\omega^4}{F} \right\rangle^{-1}\left\langle \omega^3 \right\rangle\\
a_{2}&=\frac{2g}{N}\left\langle \frac{\omega^4}{F} \right\rangle \left\langle \frac{\omega^5}{F} \right\rangle^{-1}
\end{split}
\te 
The transport coefficients $a_1$ and $a_2$ are novel coefficients that couple the nonequilibrium tensor $\gamma^{\rho\sigma}$ to color degrees of freedom. A nonvanishing $a_1$ implies that the varying color chemical potentials affect  the evolution of $\gamma^{\rho\sigma}$. The term containing $a_2$ represents the coupling of $\gamma^{\rho\sigma}$ to the gauge fields. 

If $\tau_\pi \rightarrow 0$ and $\zeta^{a(1)}=0$ in Eq. (\ref{finaleom}), we recover the (colorless) Navier-Stokes limit with $\gamma^{\mu\nu} \rightarrow -\eta \sigma^{\mu\nu}$. On the other hand, we have already shown in \cite{dev} that if we expand $\gamma^{\mu\nu}$ to second order in velocity gradients, the above formalism (with $\zeta^{a(1)}=0$) goes over to the second order conformal hydrodynamics that was derived in Refs. \cite{hyd1,hyd2}.

The equations of motion presented in this section are the main result of this work. The essential features of the effective formalism developed here are that it is nonlinear and that it is not tied up in any way to a gradient expansion. Note that the nonequilibrium tensor $\gamma^{\mu\nu}$, from which the shear tensor $\Pi^{\mu\nu}$ is obtained {\it a posteriori} as a quadratic function, satisfies a differential equation (\ref{finaleom}) instead of being an algebraic function of  velocity gradients (as it is $\Pi^{\mu\nu}$ in fluid dynamics). In the colorless case, it was shown in \cite{dev,app} that this feature results in a faster isotropization of the pressure as compared to second order hydrodynamics. Moreover, as opposed to the case of hydrodynamics, in the (colorless) effective theory the longitudinal pressure is positive throughout the entire evolution. We note that similar results were obtained in \cite{mauricio} within the so-called anisotropic hydrodynamics approach. We expect that these two results hold also in the present case including color degrees of freedom, although numerical simulations are needed to verify this. 

The color fugacities enter nontrivially in both sides of the total stress-energy tensor conservation equation, Eq. (\ref{eomtfinal}). In the left-hand side they enter through the expression for the matter stress-energy tensor, Eq. (\ref{pi2}), while in the right-hand side they enter through the coupling to nonabelian fields. Moreover, both the hydrodynamic variables $(u^\mu, \rho)$ and the nonhydrodynamic variable $\gamma^{\mu\nu}$ couple to the color fields and to $\zeta^{a(1)}$ in the evolution equation (\ref{finaleom}). 

From Eq. (\ref{finaleom}), it is seen that the dynamics of the system is highly nontrivial. In part, this is because the color chemical potentials must adjust to the flow as well as to the evolving gauge fields in order to make the system globally colorless (recall that we obtained the evolution equation from the singlet sector of the kinetic equation). We emphasize that a nonvanishing color current does not imply that the system as a whole carries a finite color charge, because the space-time dependence of the color chemical potentials can be such that the total color charge vanishes \cite{ManMro}.

The information about the constituents of the microscopic theory is encoded in the transport coefficients of the effective theory, and should in principle be computed from the former. However, the transport coefficients can also be treated as adjustable parameters. A well-known example is the case of fluid dynamics, which is usually derived from kinetic theory in the weakly coupled limit and under the relaxation time approximation (to first order in $\tau$), but then used to describe strongly coupled matter \cite{revhydro1,revhydro2}. This is done by replacing the transport coefficients of the kinetic theory by those corresponding to strong coupling, which must be computed by different means, for example, by using the AdS/CFT correspondence \cite{hyd1,hyd2}. A similar route can be taken with the various transport coefficients that arise in our formalism (at linear order in $\gamma^{ij}$ these are $(\eta,\tau_\pi,a_1,a_2)$). 
In this regard, it is important to emphasize that the effective theory presented here is consistent {\it by itself}, independently of its derivation from kinetic theory (this is also true in the case of second order hydrodynamics \cite{hyd1,hyd2}), because it satisfies the Second Law and it is expected to be causal (the colorless version was shown to be causal in \cite{dev}).

The reason for considering the colorless part of the transport equation is that, as discussed in \cite{Bod98,white}, the color currents can persist in the plasma significantly longer than the color charge density, which is neutralized rather fast. Therefore, a reasonable hypothesis is to assume that the dynamics of the system is determined by the singlet part of the distribution function. What this means is that, even though the particles carry color (and thus interact with nonabelian fields), what we are actually describing with the effective theory is the collective (or macroscopic) behavior of these particles, and it is this collective flow that is colorless. This hypothesis is physically well-motivated because one does not expect the plasma to be globally colorful, and due to this fact it has been adopted in previous studies dealing with chromohydrodynamics \cite{ManMro,manna,jiang,wakejiang} and kinetic theory \cite{mannakin,sch-WYM,sch} (see also \cite{Heinz} for a related discussion of this issue in the context of the so-called ``color hierarchy'' transport equations).

\subsubsection{Yang-Mills equation}

In order to obtain a selfconsistent system of equations for the variables $(\rho,u^\mu,\zeta^{a(1)},\gamma^{\mu\nu}, A^{a\mu})$, the conservation equations (\ref{eomtfinal}) and (\ref{eomzeta1}) and the evolution equation (\ref{finaleom}) must be supplemented with Yang-Mills equation for the gauge fields. For completeness, we write it down  
\be 
\begin{split}
&(A^{b\nu})^{;\mu}_{;\mu} - (A^{b\mu})^{;\nu}_{;\mu} + gC_{cd}^b (A^{c\mu}A^{d\nu})_{;\mu} 
+ g A_\mu^e C_{ef}^b F^{f\mu\nu} \\ 
& = \bar{n} u^\nu \bigg( \zeta^{b(1)} + \frac{1}{4}d^b_{ad}\zeta^{a(1)} \zeta^{d(1)} \bigg)
\label{ymsplit}
\end{split}
\te

\section{Relation to matrix chromohydrodynamics}
\label{matrix}
In this section we will compare the linearized versions of our approach
to linearized matrix chromohydrodynamics of Ref. \cite{ManMro}. We remark
that beyond the linear order it is not possible to establish a simple
mapping between the approach of Ref. \cite{ManMro}, which relies upon a gradient
expansion, and the one presented here, which does not.

Although our approach involving nonhydrodynamic variables is not tied up to a gradient expansion, for a local equilibrium state $\lambda^{(1)}_{ij}$ vanishes and so the effective theory reduces to ideal fluid dynamics. 
The ideal fluid chromohydrodynamic approach has been discussed in \cite{Heinz,ManMro,holm}, and later on applied to diverse studies mostly related to plasma instabilities and medium-jet interaction in the context of heavy ion collisions \cite{localeq,manna,jiang,wakejiang}. 

In the matrix chromohydrodynamic approach \cite{ManMro}, the standard procedure is to linearize the matrix equations 
\be
\mathbf{D}_\nu \mathbf{T}^{\mu\nu}=\mathbf{J}_\nu \mathbf{F}^{\nu \mu}
\te
\be 
\mathbf{D}_\mu \mathbf{J}^\mu=0
\te
together with Yang-Mills equation in matrix fluctuations $\delta \mathbf{u}^\mu,\delta \mathbf{\rho}, \delta J^a_{\nu},\delta F^{a\nu\mu}$ with respect to a given background. Aditionally, a relation between fluctuations of the matrix energy density $\mathbf{\rho}$ and the matrix pressure $\mathbf{p}$ is used. To the best of our knowledge, in previous studies based on the matrix approach the relation used in all cases is $(\delta p)_a=c_s^2(\delta \rho)_a$. It is worth noting that the use of $\delta p = c_s^2\delta \rho$ in our approach is  completely equivalent to the one adopted in the matrix formalism, as will be shown shortly. In studies dealing with chromohydrodynamics, usually the ideal fluid case is considered; for a recent extension to the Navier-Stokes case see \cite{jiang,wakejiang}. 

\subsection{Generalities}

To understand the connection between ideal or Navier-Stokes chromohydrodynamics and our effective theory, we first note that our development of the effective theory was ultimately based on Eq. (\ref{8}), which determines the coupling of matter to fields. In Eq.  (\ref{8}), the gauge fields are coupled to the {\it trace} of the matter energy-momentum tensor. 
Therefore, we end up with evolution equations for $\rm{tr}(\mathbf{T}^{\mu\nu})$, and not for $\mathbf{T}^{\mu\nu}$ itself, coupled to the Yang-Mills equation. Instead, the chromohydrodynamics of \cite{ManMro} is based on $\mathbf{T}^{\mu\nu}$, which requires the introduction of color {\it matrices} $\mathbf{u}^\mu,\mathbf{\rho}$. 

The color current, defined in Eq. (\ref{10}), is then written as $\mathbf{J}^\mu=\mathbf{n}\mathbf{u}^\mu$. 
In our formalism, the quantity $\bar{n}\zeta^a$ can be interpreted (at least at linear order) as the {\it average} color charge of a stream of colored classical particles. This interpretation for $\zeta^a$ can be seen from its {\it linearized} equation of motion given in Eq. (\ref{eomzeta1})
\be 
\bar{n}\dot{\zeta}^{f(1)} 
+ g\bar{n} u^\mu C^f_{ab} A^a_{\mu} \zeta^{b(1)} = 0
\te 
We see that it is identical to Wong's equation \cite{wong} (see also \cite{libro,winter,Selikhov,Litim}) for an {\it average} classical color charge $Q^a \equiv \bar{n}\zeta^{a(1)}$, with the time derivative of the particle's trajectory replaced by $u^\mu$, i.e., the flow velocity.  

Having the interpretation for $\zeta_a$ described above in mind, the linearized colored  fluctuations of the velocity and the four-flow $\mathbf{n}$ of the matrix approach can be written in terms of the scalar fluctuations used in our approach as $\delta u^{a\mu} = \bar{\zeta}^{a(1)}\delta u^\mu$ and $\delta n^a = \bar{n}\zeta^{a(1)}$, where $\bar{\zeta}^{a(1)}$ stands for the background value, which must be nonzero to avoid ending up in the case of a truly colorless system, as opposed to one which {\it has} color degrees of freedom but is in a colorless equilibrium state. This mapping will be used in the next section to obtain the polarization tensor of the colored plasma.

\subsection{A simple example: The polarization tensor}
\label{polar}

The polarization tensor characterizes the linear response of the system to external perturbations \cite{libro,revBlaizot,Litim,ichi,krall,fluct,Heinz,romstrick,mro89}, and it is therefore an interesting quantity to compute in the formalism presented here. With the mapping between the matrix and our approach described in the previous section, it is straightforward to show that indeed the linearized equations of motion have the same structure, and thus a very similar polarization tensor is obtained. However, the linearized equations are the same only if we set $\gamma^{\mu\nu}=0$ or $\gamma^{\mu\nu}=-\eta \sigma^{\mu\nu}$, which were the only cases studied so far \cite{ManMro,manna,jiang}. 

Our formalism naturally incorporates higher order velocity gradients, since the dissipative tensor $\gamma^{\mu\nu}$ evolves according to a differential equation rather than being expressed as an algebraic function of velocity gradients, e.g. at first order as $\gamma^{\mu\nu}=-\eta \sigma^{\mu\nu}$. It is therefore interesting to investigate the role of higher order terms on the polarization tensor. Such terms are important, for instance, in the earliest stage of evolution of the QGP created in heavy ion collisions or in situations where a fast parton goes through the QGP \cite{neuf,neufmach}. 

In order to quantify the impact of higher order viscous terms on the linear response of the system to an external gauge field, we compute the polarization tensor explicitly in our setting. The polarization tensor is defined through
\be 
\delta J_a^\mu=-\Gamma^{\mu\nu}_{ab}A_{b\nu}
\te 
where $A_{b\nu}$ is a small external perturbation. For simplicity, we shall consider an homogeneous, stationary and colorless background described by $\bar{n}$, $\bar{u}^\mu$ and $\bar{\rho}$. 
Using the mapping described above, namely $\delta u^{a\mu} = \bar{\zeta}^{a(1)}\delta u^\mu$ and $\delta n^a = \bar{n}\zeta^{a(1)}$, together with $(\delta \rho)^a \equiv \bar{\zeta}^{a(1)}\delta \rho$ and $\gamma^{a\mu\nu} \equiv \bar{\zeta}^{a(1)} \gamma^{\mu\nu}$, our linearized equations become (covariant derivatives become ordinary derivatives at this order)
\be 
\begin{split}
&\bar{n}\partial_\mu \delta u^\mu_a + \bar{u}^\mu \partial_\mu \delta n_a = 0 \\
&\bar{u}^\mu \partial_\mu \delta \rho_a + (1+c_s^2)\bar{\rho}\partial_\mu \delta u^\mu_a = 0 \\
&c_s^2 \bar{\Delta}^{\mu\nu}\partial_\nu \delta \rho_a + 
(1+c_s^2)\bar{\rho}\bar{u}^\alpha \partial_\alpha \delta u^\mu_a \\
& - \bar{n}\bar{u}_\alpha F^{\mu\alpha}_a + 
\partial_\alpha \gamma^{\mu\alpha}_a = 0 \\
&\bar{u}^\mu \partial_\mu \gamma^{\delta \rho}_a = -\frac{1}{\tau_\pi}(\eta \sigma^{\delta\rho}_a + \gamma^{\delta\rho}_a) \\
&\partial_\nu F^{\mu\nu}_a = \bar{u}^\mu \delta n_a + \bar{n}\delta u^\mu_a
\end{split}
\label{lineari}
\te 
where we have put $\bar{\Delta}^{\mu\nu}=g^{\mu\nu}+\bar{u}^\mu \bar{u}^\nu$. 
For reasons that will become clear soon, in the above equations we use a generic squared speed of sound $c_s^2$ instead of the conformal value $c_s^2=1/3$, i.e. the relation between pressure and energy density perturbations reads $\delta p = c_s^2 \delta \rho$. 

Performing a Fourier transformation we can express $\delta J_a^\mu$ in terms of the gauge field perturbation and thus find the polarization tensor (the calculation is very similar to that presented in Ref. \cite{jiang}). We get
\be
\begin{split}
\Gamma_{ab}^{\mu\nu} &= -\delta_{ab} \bigg( \frac{\omega_{\rm{pl}}^2}{(1+W_2 W_4)(k\cdot \bar{u})^2} \times \\
& \bigg[ W_5 - k^2\bar{u}^\mu \bar{u}^\nu -(k\cdot \bar{u})^2 g^{\mu\nu} \\
& +(W_1+W_3)[k^2 W_5 - k^\mu k^\nu (k\cdot \bar{u})^2 - 
k^4\bar{u}^\mu \bar{u}^\nu] \bigg]  \bigg)
\end{split}
\te
where 
\be
\begin{split}
W_1 &= -\bigg(k^2+(c_s^{-2}-1)(k\cdot \bar{u}) \bigg)^{-1} \\
W_2 &= \frac{\eta}{(1+c_s^2)\bar{\rho}(k\cdot \bar{u})[1+i\tau_\pi (k\cdot \bar{u})]} \\
W_3 &= -\frac{W_2(1+4W_1 W_4)}{3+3c_s^2\frac{W_4}{(k\cdot \bar{u})^2}+4W_2 W_4} \\
W_4 &= k^2 - (k\cdot \bar{u})^2 \\
W_5^{\mu\nu} &= (k\cdot \bar{u})(\bar{u}^\mu k^\nu + k^\mu \bar{u}^\nu)
\end{split}
\te 
and 
\be 
\omega_{\rm{pl}}^2 = \frac{\bar{n}^2}{(1+c_s^2)\bar{\rho}} 
\te
is the plasma frequency. Note that $\Gamma_{ab}^{\mu\nu}$ is diagonal in color space, as expected since as shown by Eqs. (\ref{lineari}) there is no mixing of colors at linear order, and transverse with respect to $k^\mu$. 

The result for $\Gamma_{ab}^{\mu\nu}$ that we obtain is the same as that obtained in Ref. \cite{jiang} but with the shear viscosity $\eta$ replaced by an effective one 
\be 
\eta_{\rm{eff}}=\frac{\eta}{1+i\tau_\pi (k\cdot \bar{u})}
\te
The appearance of $\eta_{\rm{eff}}$ in place of $\eta$ is quite natural since $\tau_\pi$ is precisely the relaxation time of the shear tensor $\Pi^{\mu\nu}$ towards its Navier-Stokes value. We emphasize that this similarity with the chromohydrodynamic result holds only when linearizing the equations. The fully nonlinear evolution equations of the effective theory developed here, which do not involve hydrodynamic gradients, are different from those of Navier-Stokes chromohydrodynamics. 

To quantify the effect of higher order viscous terms on the linear response of the plasma to an external gauge field, we will work with the longitudinal part $\epsilon_L$ of the dielectric tensor $\epsilon^{ij}$. Similar analysis can be performed for the transverse part of the dielectric tensor, $\epsilon_T$,  but for brevity we shall only consider $\epsilon_L$. We have (we suppress color indices and put $k^\mu=(\omega,\mathbf{k})$ in what follows)
\be 
\epsilon^{ij}= \delta^{ij} + \frac{1}{\omega^2} \Gamma^{ij}
\te
and 
\be 
\epsilon_L = \frac{k_i k_j \epsilon^{ij}}{\mathbf{k}^2}
\te 
In the rest frame $\bar{u}^\mu=(1,0,0,0)$ we get 
\be
\epsilon_L(\omega,k) = 1-\frac{\omega_{\rm{pl}}^2}{\omega^2}\bigg(\frac{1}{1-W_2 \mathbf{k}^2}\bigg)(1-(W_1+W_3))
\te 
where we can use that $(1+c_s^2)\bar{\rho}=sT$ ($\bar{s}$ is the entropy density) to rewrite $W_2$ as 
\be
W_2 = \frac{\eta}{sT \omega (1+i\tau_\pi \omega)} 
\te

For illustrative purposes, and following \cite{jiang}, we focus on the soft modes $\omega,\sqrt{\mathbf{k}^2} \ll T$ and set $\sqrt{\mathbf{k}^2}=2 \omega_{\rm{pl}}$ and $T=10\omega_{\rm{pl}}$. The relaxation time is set to its value computed from the kinetic theory of a Boltzmann gas (without color), which is given by  $\tau_\pi = 6 \eta/(sT)$. As a typical value for the temperature we shall use $T=200$ MeV. 

The comparison made between ideal chromohydrodynamics and kinetic theory carried out in Ref. \cite{mannakin} in the context of jet-induced instabilities shows that in order to achieve reasonable agreement between both descriptions an effective speed of sound must be used in the former. 
To be consistent with previous studies we will show numerical results obtained with this effective speed of sound (instead of $c_s^2=1/3$), which is given by 
\be 
c_s^2 = \frac{1}{3}\bigg[ 1+\frac{1}{2y}\ln \bigg(\frac{1-y}{1+y}\bigg) \bigg]^{-1} + \frac{1}{y^2}
\label{ceff}
\te 
where $y=\sqrt{\mathbf{k}^2}/\omega$. We note that this expression for $c_s^2$ is chosen to make the longitudinal dielectric function obtained from the ideal  chromohydrodynamic approach of Ref. \cite{mannakin} identical to that obtained from kinetic theory in the leading-order HTL approximation  \cite{libro,reviewIANCU,revBlaizot,BraPis90a,FreTay90,Bod98,Bod99,ArSoYa99a,ArSoYa99b,Litim}. Specifically, in the soft limit $\omega,\sqrt{\mathbf{k}^2} \ll T$ and putting $\mathbf{k}=(k,0,0)$ for simplicity we have \cite{jiang} 

\be 
\begin{split}
\epsilon_L &= 1+\frac{3\omega_{\rm{pl}}^2}{k^2}\bigg( 1-\frac{\omega}{2k} 
\bigg[ \ln \bigg| \frac{\omega+k}{\omega-k}\bigg| - i\pi \theta(k^2-\omega^2)\bigg] \bigg) \\
&- \frac{12\omega_{\rm{pl}}^2}{k^2}\frac{\eta_{\rm{eff}} \omega}{sT} 
\bigg( 1-\frac{\omega}{k} \ln \bigg| \frac{\omega+k}{\omega-k} \bigg| 
+ \frac{\omega^2}{4k^2}\bigg[ \ln \bigg|\frac{\omega+k}{\omega-k}\bigg| \bigg]^2 \\
&-\frac{\omega^2}{4k^2}\pi^2\theta(k^2-\omega^2) \\
&+ i \pi \bigg[\frac{\omega}{k} - 
\frac{\omega^2}{2k^2}\ln \bigg| \frac{\omega+k}{\omega-k} \bigg|\bigg]\theta(k^2-\omega^2) 
\bigg)
\end{split}
\label{elfull}
\te 
where $\delta \rightarrow 0^+$ and we used that 
\be
\ln \bigg[\frac{\omega + k + i\delta}{\omega - k + i\delta} \bigg] = \ln \bigg| \frac{\omega+k}{\omega-k}  \bigg| - i\pi \theta(k^2-\omega^2) 
\te
with $\theta(x)$ the Heavyside step function. 
This shows that, in the soft limit, the modes with $\omega>k$ are undamped if $\tau_\pi = 0$. If the relaxation time does not vanish, the imaginary part of $\epsilon_L$ is nonzero even for $\omega>k$. We will now show some illustrative examples of this feature. 

\begin{figure}[htb]
\begin{center}
\scalebox{0.46}{\includegraphics{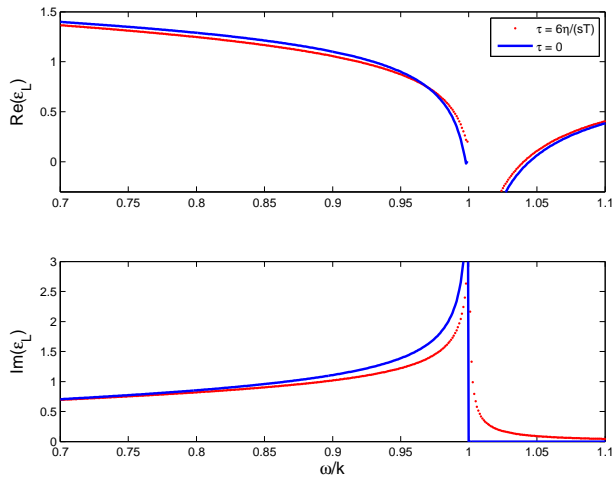}}
\caption{(Color online) Real (upper panel) and imaginary (lower panel) parts of the longitudinal dielectric function $\epsilon_L$ as a function of frequency, for $\eta/s=0.3$. The values of the parameters are set to $k=2 \omega_{\rm{pl}}$, $T=10\omega_{\rm{pl}}$, $\tau_\pi = 6 \eta/(sT)$ or $\tau_\pi=0$, and $T=200$ MeV. The results are obtained using the effective speed of sound given by Eq. (\ref{ceff}).}
\label{p03}
\end{center}
\end{figure}

Figure \ref{p03} shows the real and imaginary parts of $\epsilon_L$ as a function of $\omega/k$ for $\eta/s = 0.3$. This value for $\eta/s$ is on the high side in terms of fitting viscous fluid dynamics results to RHIC and LHC data \cite{revhydro1,revhydro2,reviewIANCU}. It is seen that there are significant differences between the longitudinal dielectric function computed with different values of the relaxation time. The most noticeable effects are seen on the imaginary part of $\epsilon_L$, which, for $\omega < k$, is smaller in the case with nonvanishing $\tau_\pi$. This indicates that, in this range of frequencies, the induced color excitations decay more slowly as compared to the case with $\tau_\pi=0$. This behavior can be understood by recalling the physical meaning of $\tau_\pi$ as the relaxation time of the shear tensor $\Pi^{\mu\nu}$ towards its Navier-Stokes value $-\eta \sigma^{\mu\nu}$. If the value of $\tau_\pi$ is increased, then hydrodynamic fluctuations will decay more slowly. Since hydrodynamic fluctuations are coupled to color fluctuations, the latter will decay more slowly as well. This result is in agreement with those of Ref. \cite{sch} obtained from kinetic theory with a BGK collision kernel, showing that the addition of hard-particle collisions slows the rate of growth of QCD plasma unstable modes.  

As shown in Figure \ref{p03}, the results that we obtain for $\omega > k$ show that the damping of color excitations in this frequency range is completely different according to whether $\tau_\pi$ is zero or not. As expected, if $\tau_\pi=0$ there is no damping in this frequency range. This feature stems from the analytic structure of the longitudinal dielectric function in the regime where $\omega, k \ll T$, as given by Eq. (\ref{elfull}). In this limit, the imaginary part of $\epsilon_L$ is proportional to the step function $\theta(k^2-\omega^2)$ (see e.g. \cite{Litim}). 
On the contrary, if $\tau_\pi \neq 0$ then those color excitations with $\omega \gtrsim k$ become  considerably damped, with a damping rate which falls off steeply with increasing frequency. Similar results were obtained in Ref. \cite{carrdiel} for QED dispersion relations obtained from kinetic theory with a BGK collision term. There it was found that when collisions are included, the longitudinal dispersion intersects the light cone $\omega=k$, in contrast to the case of collisionless dispersion where $\omega > k$ for all $k$. In a collisionless plasma, Landau damping, which is possible only for $\omega < k$, is the only damping mechanism, and thus plasma waves are undamped. In contrast, collisions introduce an additional damping mechanism for plasma waves (see e.g. \cite{sch}). We emphasize that a first order hydrodynamic formalism, in which the shear tensor relaxes instantaneously to its Navier-Stokes value, can not completely account for such damping of plasma waves. 

\begin{figure}[htb]
\begin{center}
\scalebox{0.46}{\includegraphics{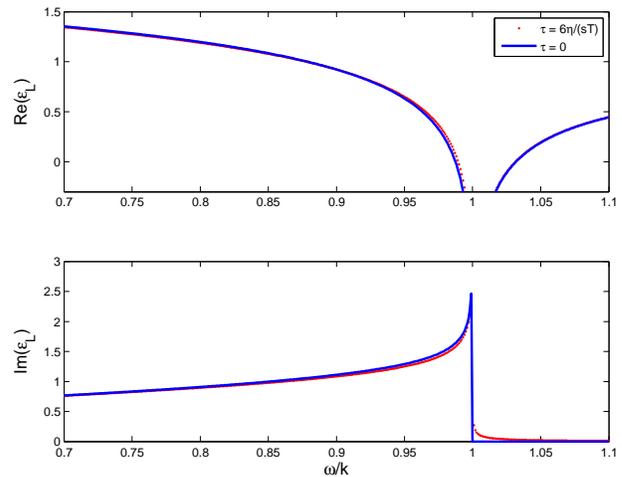}}
\caption{(Color online) Real (upper panel) and imaginary (lower panel) parts of the longitudinal dielectric function $\epsilon_L$ as a function of frequency, for $\eta/s=0.15$. The values of the parameters are set to $k=2 \omega_{\rm{pl}}$, $T=10\omega_{\rm{pl}}$, $\tau_\pi = 6 \eta/(sT)$ or $\tau_\pi=0$, and $T=200$ MeV. The results are obtained using the effective speed of sound given by Eq. (\ref{ceff}).}
\label{p015}
\end{center}
\end{figure}

We now briefly discuss the influence of the value of $\eta/s$ on the longitudinal dielectric function. Figure \ref{p015} shows the real and imaginary parts of $\epsilon_L$ for a smaller value of the viscosity-to-entropy ratio than before, namely $\eta/s = 0.15$. It is seen that the impact of a nonvanishing relaxation time on $\epsilon_L$ decreases with decreasing values of $\eta/s$. For the value  $\eta/s = 0.15$, the effect of $\tau_\pi$ on $\epsilon_L$ is still significant, and we still see that excitations with $\omega \gtrsim k$ are damped due to collisions. 
Although not shown, we find that for $\eta/s \lesssim 0.08$ the difference between the longitudinal dielectric function obtained with a vanishing or a nonvanishing value of $\tau_\pi$ is hardly appreciable.  

The effective formalism discussed in this work is phenomenological and involves several approximations. In spite of this, it is interesting to  qualitatively discuss possible implications of our results for the phenomenon of jet quenching in heavy ion collisions. For this, and considering the already discussed limitations, we take the view that our formalism constitutes an appropriate model to understand some features of the response of the QGP to a fast moving parton that crosses it. 

There are two main energy loss mechanisms which contribute to energy loss: radiation of soft gluons and collisions involving the exchange of {\it hard} ($\sim T$ or larger) and
{\it soft} ($\sim 2\omega_{\rm{pl}}$) momenta \cite{loss1,loss2}. The dominant source of energy loss is gluon bremsstrahlung, although the contribution of collisions to the total energy loss is significant, particularly when attempting to fit the results of theoretical models of jet quenching to data \cite{loss1,loss2,losscomp,losscomp2}. In connection to our results, we note that the contribution of soft collisions to the energy loss can be directly calculated 
from $\epsilon_L$ and $\epsilon_T$ (see for example \cite{ichi,bluhm,mullerloss}). We shall not discuss in detail this issue here, but just mention that a smaller imaginary part of $\epsilon_L$ will result in a decrease in the energy loss. Our results then show that a nonvanishing value of the relaxation time $\tau_\pi$ is expected to lead to a  sizeable reduction in the energy loss. 

As a final remark, we note that the radiation spectrum and hence the energy loss of a hard parton crossing the QGP is also modified by the dielectric polarization of the medium (this is known as the Ter-Mikaelian effect \cite{TMorig} - see \cite{TM} for an extension to QCD). As noted recently \cite{bluhm}, the effect of radiation damping occurring in an absorptive medium on the spectrum of radiated gluons is particularly interesting and might lead to sizeable effects on energy loss related phenomena. The polarization tensor derived from the formalism presented here naturally incorporates damping. A detailed study of the influence of $\tau_\pi$ on the energy loss of fast partons for the conditions prevailing in heavy ion collisions at RHIC and LHC is left for future work. We emphasize, however, that the richness of the effective formalism presented here lies in its nonlinear character, which is not reflected in the polarization tensor but may become relevant when dealing with parton energy loss phenomena .

\section{Summary and outlook}
\label{summ}

In this work we have obtained from the kinetic theory of nonabelian plasmas an effective model describing the evolution of a system composed of colored particles interacting with nonabelian classical gauge fields. The link between the one-particle distribution function of colored particles in the kinetic description and the variables of the effective theory is determined by the entropy production variational method. The closure provided by this method does not rely in any way on a gradient expansion in macroscopic variables and can therefore be applied even when these gradients are large.

In order to compare the developed effective theory with chromohydrodynamic formalisms based on the usual gradient expansion, we have calculated the longitudinal dielectric function $\epsilon_L$ of the plasma. Using typical values of the plasma parameters appropriate for the QGP, together with an effective speed of sound chosen to reproduce the longitudinal dielectric function of hard-thermal loop kinetic theory, we have found that the relaxation time for the shear tensor has a strong influence on the dynamics of color fluctuations, in agreement with the results of kinetic theory including collisions among the hard partons.  The implications of such changes on the evolution of color excitations on phenomena relevant to heavy ion collisions, particularly on jet quenching, deserve further investigation.

The formalism presented here is a simplified model of the dynamics of color fields during the early and intermediate stages of heavy ion collisions. It may be useful to shed light on issues which would require intensive simulations in a microscopic approach, for example the magnitude of the back reaction of the particles' flow on the gauge fields. If the color fields eventually die out, the effective theory goes over to second order fluid dynamics if the velocity gradients are small, so that the effective theory could be used to describe (starting from suitable initial conditions) the evolution of the fireball created in a heavy ion collision from very early times ($\gtrsim 0.2$ fm/c) till freeze-out in a unified, albeit simplified, way. 

Concerning the dynamics of color fields at early and intermediate times, our formalism could be used to study plasma instabilities and its effect on the evolution of matter created in heavy ion collisions. It would be particularly interesting to solve numerically the full {\it nonlinear}  equations of the effective theory presented here for the conditions prevailing in heavy ion collisions, and to compare the results to those obtained by a microscopic approach. To carry out this program, the inclusion of hard gluons into the model is certainly required for a realistic description of the physical processes involved at those stages. Work is in progress along these lines.

At this point we would like to comment on the limitations of our approach in connection to possible applications to describe the early-time dynamics of color fields in heavy ion collisions (some of the issues discussed here are also relevant for parton energy loss). We think that it is clearer to distinguish between limitations inherent to our approach (that are either truly unsurmountable or else very difficult to address) and simplifying hypothesis that could be relaxed in the future. 

We start by discussing those limitations that are inherent to our approach. 

The effective theory presented here constitutes a simplified model of the true dynamics given by kinetic theory with a linear collision term, and therefore the equations derived in this paper should be applicable for similar time scales.  
The kinetic theory description of the early stage of heavy ion collisions is valid for times $\gtrsim Q_s^{-1}$ ($\sim 0.2$ fm/c at RHIC) when
particles having transverse momenta greater than $Q_s$ are formed out of the color fields \cite{reviewIANCU,GLASMA}. 
However, there are some limiting factors that should be considered. 

The first one was already hinted to and refers to the use of a linear collision term. Although one could, in principle, write down the variational equations of the EPP for a transport equation with nonlinear collision kernels, there is not much prospect of being able to solve them. However, this may not be a serious issue, because several studies have shown that the Boltzmann-Vlasov equation coupled to Yang-Mills equation provides a fairly reliable description of the early stages of a heavy ion collision \cite{mro93,mro89,mro94,mro97,strick-Weibel,arn-Weibel,mannakin,fluct,strick,rom1,rom2,muller-pos,flor,ipp}. The short-range interaction between hard partons has been incorporated in the kinetic models only recently \cite{sch,sch-WYM,dum,nayak,Gale}.  

The second point is that the range of applicability of the effective theory developed here is not determined by the magnitude of velocity gradients, but rather by whether the dynamics of the system as given by (linearized) kinetic theory can or can not be described by few variables (including nonhydrodynamic ones) coupled to classical gauge fields. As it happens with other approaches to the closure problem \cite{deGroot,denicol,ichi,krall,GKbook,GKpaper,anile,tanos,tanos2,tanos3,muscato,christen,christen2,ferz,nagy,geroch,liboff}, it is difficult to precisely establish {\it a priori} the range of validity of the resulting effective theory. We believe that, as it happens with fluid dynamics, our formalism may prove useful to describe some stages of a heavy ion collision provided the transport coefficients are suitably chosen. In any case, this point should be settled by comparing the results obtained from the effective theory to those obtained from kinetic theory. 

As an issue that can be improved in the future, we mention first the fact that here we deal with excitations of scalar fields (and not spinors), and second that we do not take into account the hard gluons. In the early stage of heavy ion collisions, the hard gluons are, as the hard quarks, coupled to the soft gluons and therefore should also be described by a kinetic equation with its corresponding collision term. This term would involve not only interactions among hard gluons themselves but also between hard gluons and the excitations of the scalar fields. The latter coupling should also be reflected in $\mathbf{I}_{col}$ of Eq. (\ref{9}). The inclusion of hard gluons interacting with the classical gauge fields is mandatory for the model to be applicable to the early stage of a heavy ion collision, in which strong fields decay into gluons to eventually form the QGP. 

Another point that can be addressed in the future is the possibility of performing a quadratic expansion around an anisotropic (in momentum space) distribution function instead of the isotropic distribution function we use here. This idea was exploited in \cite{mauricio,aniso} to obtain an  effective theory that can handle the very large anisotropies in momentum space that are present at early times in heavy ion collisions, and moreover reproduces both the ideal fluid and the free-streaming limits. The correct description of both regimes (which are ultimately determined by the value of the Knundsen number) may be a relevant issue when dealing with plasma instabilities. Although our formalism can handle very large anisotropies as well, at present it is not clear how well can it describe plasma instabilities, so further studies are certainly needed to settle this point. 

Finally, we plan to include stochastic terms into the evolution equations of the effective theory \cite{cfluc,nosjhep}. This would allow us to address in a simple model setup the important question on the fate of fluctuations and their impact on observables at RHIC and LHC.

\begin{acknowledgments}
We thank Gast\~ao Krein and Mauricio Martinez for useful comments and discussions. 
This work has been supported in part by ANPCyT, CONICET and UBA under project UBACYT X032 (Argentina).
\end{acknowledgments}

\end{document}